\documentclass[journal,comsoc]{IEEEtran}
\usepackage[T1]{fontenc}

\ifCLASSINFOpdf
\else
\fi
\usepackage{rotating,threeparttable,booktabs,caption,dcolumn}
\newcolumntype{d}[1]{D..{#1}}

\usepackage{subcaption}
\usepackage{amsmath,amssymb,amsfonts}
\usepackage{wrapfig, framed, caption}
\usepackage[lined,ruled,linesnumbered]{algorithm2e}
\usepackage{algorithmic}
\usepackage{graphicx}
\usepackage{url}
\usepackage{textcomp}
\usepackage{booktabs}
\usepackage{array}
\usepackage{float}
\usepackage{multirow}
 
\def\BibTeX{{\rm B\kern-.05em{\sc i\kern-.025em b}\kern-.08em
    T\kern-.1667em\lower.7ex\hbox{E}\kern-.125emX}}
\usepackage{multirow}
\usepackage{enumerate}

\usepackage[cmintegrals]{newtxmath}
\begin{document}
%
\title{Layer-Wise Partitioning and Merging for Efficient and Scalable Deep Learning}
%
%
%

        

\author{Samson B. Akintoye, Liangxiu Han, Huw Lloyd, Xin Zhang, Darren Dancey, Haoming Chen, and Daoqiang Zhang
\thanks{S. B. Akintoye, L. Han, H. Lloyd, X. Zhang, D. Dancey are with the Department of Computing and Mathematics, Manchester Metropolitan University, UK (e-mail: s.akintoye@mmu.ac.uk; l.han@mmu.ac.uk; huw.lloyd@mmu.ac.uk; x.zhang@mmu.ac.uk; d.dancey@mmu.ac.uk)}
\thanks{H. Chen is with the Department of Computer Science, University of Sheffield, UK (e-mail: hchen78@sheffield.ac.uk)}
\thanks{D. Zhang is with the College of Computer Science and Technology, Nanjing University of Aeronautics and Astronautics, P.R.China (e-mail: dqzhang@nuaa.edu.cn)}
\thanks{Corresponding author: L. Han (e-mail: l.han@mmu.ac.uk)}}

%
%

\markboth{Journal of \LaTeX\ Class Files,~Vol.~14, No.~8, August~2015}%
{Shell \MakeLowercase{\textit{et al.}}: Bare Demo of IEEEtran.cls for IEEE Communications Society Journals}
%



\maketitle

\begin{abstract}

\end{abstract}
Deep Neural Network (DNN) models are usually trained sequentially from one layer to another, which causes forward, backward and update locking's problems, leading to poor performance in terms of training time. The existing parallel strategies to mitigate these problems provide suboptimal runtime performance. In this work, we have proposed a novel layer-wise partitioning and merging, forward and backward pass parallel framework to provide better training performance. The novelty of the proposed work consists of 1) a layer-wise partition and merging model which can minimise communication overhead between devices without the memory cost of existing strategies during the training process; 2) a forward pass and backward pass parallelisation and optimisation to address the update locking problem and minimise the total training cost. The experimental evaluation on real use cases shows that the proposed method outperforms the state-of-the-art approaches in terms of training speed; and achieves almost linear speedup without compromising the accuracy performance of the non-parallel approach.




%
\IEEEpeerreviewmaketitle

\section{Introduction}

Deep Neural Networks (DNNs) have shown promise in different applications such as computer vision, Natural language processing and speech recognition. However, training a DNN remains a significant challege, which is both computational and data intensive \cite{Szegedy2016, Hu2018, 8593683}. To mitigate this problem, DNN models are usually trained in parallel across either homogenous or heterogeneous devices including CPUs and GPUs \cite{dl:Abadi} for better training performance.
One of the common distributed training methods is model parallelism \cite{dp:Dean}; model parallelism allocates disjoint subsets of a DNN model to each dedicated device \cite{Dryden2019}. This method requires data communication between computation processes to update the model in each training iteration. The backpropagation algorithm \cite{Cheng2021DerivationOT, Fan2018GeneralBA} is usually used for the updates and consists of two phases: the forward pass and backward pass. The forward pass calculates and stores intermediate variables such as outputs for a neural network from the input to the output layer. The backward pass method calculates the gradients of neural network parameters, in reverse order, from the output to the input layer. The sequential execution of forward pass and backward pass requires data communication between computation processes to update the
model in each training iteration, usually referred to as forward, backward, and update locking problems \cite{dl:Max}, which lead to inefficient training performance due to computation dependencies.\\
Several methods have been proposed to mitigate these problems. One of these methods is {\em delayed gradients}, which breaks the backward locking \cite{dl:dg2, dl:dg1}. However, this method suffers from large memory usage due to the requirement to store all the intermediate computation results. In addition, the delayed gradients method provides suboptimal performance in terms of training speed and convergence rate when the DNN model becomes deeper and larger.  Another method is {\em feature replay} \cite{Huo2018TrainingNN, Liu2020GenerativeFR}, which also breaks backward locking and provides better performance than delayed gradients in memory consumption. The main disadvantage is that feature replay has a greater computational load than delayed gradients, leading to lower training speed. Finally, {\em layer-wise parallelisation} is a method in which each network layer is parallelised individually, with the solution to a graph search problem used to optimise the layer parallelisation \cite{Jia2018ExploringHD}. However, this method still incurs communication overhead because the computations of each layer are performed on a single device while the entire model is trained on multiple devices using the data parallelism technique, which requires sharing gradients across devices consequently limiting training performance.\\
Unlike existing methods, this paper proposes a novel layer-wise partitioning and merging approach to minimise communication overhead between devices without incurring a significant memory overhead during the training process. Our proposed method applies partition and merging operations to perform computations of network layers across available multiple devices to minimise the communication overhead. In addition, we propose a forward pass and backward pass parallelisation method to address the update locking problem associated with the sequential execution of forward pass and backward pass computations. Thus, the main contributions of this paper include:
\begin{itemize}

\item We propose a novel layer-wise partitioning and merging for efficient distribution and processing of network layer computations across multiple devices. The partitioning and merging mechanism can minimise communication overhead between devices in distributed training.

\item We propose a forward pass and backward pass parallelisation method for solving locking problems, with an associated cost function formuation for optimising training performance by reducing the total training cost.

\item We apply the proposed methods to two real use cases representing different complexity and depth of the models for performance evaluation of the proposed approach.

\end{itemize}

The remaining parts of this paper are organized as follows. In Section \ref{rel}, we summarize the related research work to this paper. In Section \ref{meth}, we introduce a layer-wise partitioning and merging, forward pass and backward pass parallelisation framework. We conduct experiments to evaluate our proposed method in Section \ref{exp}, and Section \ref{pd:conc} concludes the work and highlights the future work.

\section{Related Work}\label{rel}

The increase in dataset and DNN model sizes has motivated the use of distributed training of DNN models for better performance. Existing parallel methods are usually developed based on the {\em data} and {\em model parallelism} techniques to distribute training across multiple devices. Data parallelism divides the entire training dataset into subsets of data and dispatches on multiple devices. Each device maintains a DNN model replica and its parameters. On the other hand, model parallelism splits and trains large DNN models onto multiple computation devices instead of a single device for efficient training performance \cite{Jia2019, 6853593}. In  \cite{dp:Alex2}, data parallelism was used for convolutional and pooling layers and model parallelism for densely connected layers to accelerate CNNs training performance. Wu et al \cite{dl:Wu6} adopted data parallelism to allocate the RNN model replica on each node and model parallelism for intra-node parallelisation. Although these works improve performance over either data or model parallelism, they still provide suboptimal performance and scale poorly on large datasets and multiple devices. Saguil and Akramul \cite{9039657} proposed a layer partitioning method to improve the training performance of neural network-based embedded applications in edge networks. The method partitions layers of a model into sub-models and distributes them among different devices. The method was shown to reduce the communication overhead between devices by up to 97\% with a tradeoff of 3\% in accuracy. Similarly, the works in \cite{Martins1, Martins2, Li2020, Ko2018EdgeHostPO, 234801} split neural networks layers and allotted sub-layers to multiple devices for improved training performance. Finally, Song et al. \cite{8675232} proposed layer-wise parallelism which partitions feature map tensors, kernel tensors, gradient tensors, and error tensors, subsequently optimizing the partition with the goal of minimizing the total communication for the acceleration of the DNN training.

Aside from the data and model parallelism, {\em pipeline parallelism} \cite{Zhan2019, Yang2019, Geng2019}, and {\em hybrid parallelism} \cite{Boehm1, Ono2019HybridDP, 8675232, Akintoye2021AHP} have been proposed to speed up DNN training further. Pipeline parallelism partitions model layers into stages and runs them on multiple devices. Huang et al. \cite{Huang2019} proposed {\em GPipe}, a pipeline parallelism based solution that explores the synchronous approach to train large models and optimised GPU memory usage. Narayanan et al. \cite{Deepak2019} proposed {\em PipeDream}, which uses the hybrid method of data and pipeline parallelism for asynchronous training of the DNN models. Hybrid parallelism combines the advantages of two or more types of parallelism while weakening the disadvantages of each for better performance. 

In recent times, more parallel strategies have been proposed to improve the training performance by addressing forward, backward, and update locking problems \cite{dl:Max}. Belilovsky et al. \cite{dl:Eugene} proposed a greedy algorithm based solution known as {\em Decoupled Greedy Learning} (DGL) to achieve update unlocking as well as forward unlocking. The work decoupled and parallelized the CNN layers training to achieve better convergence performance than state-of-the-art approaches. Furthermore,  Huo et al. \cite{dl:Zhouyuan} proposed a {\em Decoupled Parallel Back-propagation} (DDG, in which the DG refers to delayed gradients), which splits the network into partitions and solves the problem of backward locking by storing delayed error gradient and intermediate activations at each partition. Similarly, Zhuang et al. \cite{dl:Huiping} adopted the delayed gradients method to propose a fully decoupled training scheme (FDG). The work breaks a  neural network into several modules and trains them concurrently and asynchronously on multiple devices. In DDG and FDG, the forward pass executes sequentially. The input data flows from one device to the other and computes sequential activation order.
On the other hand, all devices except the last one store delayed error gradients and execute the backward computation after the forward computation is complete. DDG and FDG adopt the delayed gradients to split the backward pass and reduce the total computation time to $T_{f} + T_{b}/N$, where $T_{f}$, $T_{b}$, and $N$ denote forward pass time, backward pass time, and the number of devices for a mini-batch in Naive sequential method. However, the two delayed gradients based approaches incur a large memory overhead due to storing intermediate results and suffer from weight staleness and the forward locking problem \cite{dl:Max}.\\
To address these  challenges, Xu et al. \cite{9156835} proposed {\em Layer-wise Staleness} and {\em Diversely Stale Parameters} (DSP), a combination of parallel DNN model training algorithms, where `staler' information is used to update lower layer parameters. DSP overlaps both forward pass and backward pass computations to the reduce memory consumption experienced in DDG and FDG during the training process, improving training performance. However, the staleness-based methods have slow convergence, especially when the DNN models become more complex and deeper, thereby negatively impacting training performance in terms of speed for the desired accuracy \cite{DBLP:DaiZDZX19}.
 \\
To address the limitations of the aforementioned parallel methods, we propose a novel layer-wise partitioning and merging, forward pass and backward pass parallelisation approach for accelerating distributed training of the DNN models.

\section{The proposed Method}\label{meth}
In this section, we provide full details of the proposed method, which divides into two phases: layer-wise partitioning and merging; and forward pass and backward pass parallelisation.  The proposed method aims to minimise the communication overhead and address locking problems associated with the sequential execution of forward pass and backward pass computations in a distributed environment.
\subsection{Problem Statement}
In deep learning, backpropagation is an algorithm to train feedforward neural networks, divided into forward pass and backward pass phases for sequential computations of activation and error gradient, respectively. However, the sequential calculations result in backward locking and forward locking problems due to computation dependencies between network layers. In addition, there is also an update locking problem because the backward pass waits for the forward pass to finish before it starts. 

\begin{figure}[h!]
\centering
\includegraphics[scale=0.32]{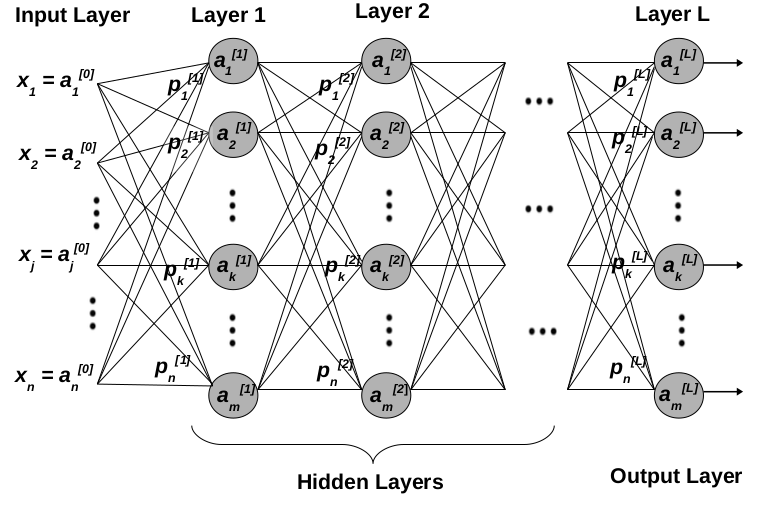} 
\caption{ Feedforward neural network \label{pd:feu}}
\end{figure}

The process of training feedforward neural networks is represented in Fig \ref{pd:feu}, and Table \ref{tb:frame} presents the notation used here for the training parameters.

\begin{table}[h!]
\begin{center}
\caption{Notation used in this paper.}
\label{tb:frame}
\begin{tabular}{{c}l*{3}{c}r{c}}
\toprule
Notations           & Descriptions   \\
\hline
$L$ & Number of network layers. \\
${p^{l}}$ & The weight parameter of $l$ layer.\\
$a^{l}$ & The activation of layer  $l$.\\
$X = a^{[0]}$ & The input data.\\
$e$ & The input-label.\\
$b$  & The Batch size. \\ 
$f$ & The Loss function.\\ 
$n$ & Number of GPUs.\\
$G$ & The activation function.\\
$\omega^{t}$ & gradient of the function at iteration $t$.\\
$T_{F}$ & forward pass time.\\
$T_{B}$ & backward pass time.\\

 \bottomrule
\end{tabular}
\end{center}
\end{table} 
We assume that a DNN model has $L$ consecutive layers and $p = (p^{[1]},p^{[2]},\dots, p^{[l]}) \in {\mathbb{R}^n}$ where $p^{[l]} \in {\mathbb{R}^n_{l}}$ denotes the weight parameter at layer $l \in \lbrace 1, 2,.., L\rbrace $ and $n =  \sum_{l=1}^{L} 2^{-n} n_{l}$. The activation of each layer  $l$ is defined as:
\begin{equation}\label{eqn1}
 a^{[l+1]} = g_{l}(a^{[l]},p^{[l]}) 
\end{equation} 
 where $a^{[l]}$ and $a^{[0]}$  are the input of layer $l$ and input data respectively. Generally, the layer's activation value can be defined as: 

\begin{equation}
\label{eq1}
\begin{split}
a^{[l+1]} & := G(a^{[1]},p^{[1]},p^{[2]},...,p^{[l]}) = \\ & g_{l}(... g_{2}( g_{1}( a^{[1]},p^{[1]}),p^{[2]}...,p^{[l]}) 
\end{split}
\end{equation}

The loss function is $f(a^{[L]},e)$, where $e$ denotes input-label detail of the training samples. The loss function of the feedforward neural network can be represented as the following optimization problem:

\begin{equation}
\label{eq2}
\min_{f} g(p) = f(G(a^{[1]},p^{[1]},p^{[2]},...,p^{[l]}), e)
\end{equation}

Gradient descent is used to solve the optimization problem given in equation~\ref{eq2} by iteratively moving in the direction of the negative of the gradient of the function at iteration $t$ is defined as: 
\begin{equation}
\label{eq3} 
\omega^{t}_{p} = [\omega^{t}_{p^{[1]}},\omega^{t}_{p^{[2]}},...,\omega^{t}_{p^{[l]}}]
\end{equation}
where,
\begin{equation}  
\omega^{t}_{p^{[l]}} = \dfrac{\delta g(p_{t})}{\delta p_{t}^{[l]}}
\end{equation}
Typically, either stochastic gradient descent (SGD) \cite{Gower2019SGDGA} or more recent algorithms such as ADAM \cite{Kingma2015AdamAM} are used to update the model parameters $p$ iteratively as:

\begin{equation}\label{eqn2} 
p_{t+1}^{[l]} = p_{t}^{[l]} - \alpha_{t} \omega^{t}_{p^{[l]}}
\end{equation} 
where $\alpha_{t}$ is the learning rate. When the training sample is large and $b_{t}$ is mini-batch of $b$, the gradient vector becomes:

\begin{equation}
\omega^{t}_{p^{[l]}} = \dfrac{\delta g_{b_{t}}(p_{t})}{\delta p_{t}^{[l]}}
\end{equation}
The backpropagation algorithm is usually used to calculate the model gradients, which consists of two processes: the forward pass for model prediction and the backward pass for gradient calculation and model update. In the backpropagation process, the input of each layer relies on the output of the immediate previous layer. For instance, the gradient in layer $l$ using the gradient back-propagated from layer $u$ and $v$ such that $l < u < v$ can be expressed as:
\begin{equation}
\omega^{t}_{p^{[l]}} = \dfrac{\delta g_{b_{t}}(p_{t})}{\delta p_{t}^{[l]}} = \frac{\delta q_{t}^{[u]}}{\delta p_{t}^{[l]}} \dfrac{\delta g_{b_{t}}(p_{t})}{\delta q_{t}^{[u]}} = \frac{\delta q_{t}^{[u]}}{\delta p_{t}^{[l]}} \omega^{t}_{q^{[u]}}
\end{equation}

where,
\begin{equation}
\omega^{t}_{q^{[u]}} = \frac{\delta g_{b_{t}}(p_{t})}{\delta q_{t}^{[u]}} = \frac{\delta q_{t}^{[v]}}{\delta q_{t}^{[u]}} \dfrac{\delta g_{b_{t}}(p_{t})}{\delta q_{t}^{[v]}} = \dfrac{\delta q_{t}^{[v]}}{\delta q_{t}^{[u]}} \omega^{t}_{q^{[v]}}
\end{equation}
Moreover, the backward process waits until the forward process is complete. This situation, often referred to as the forward, backward and update lockings problem, causes delays in the model updates, leading to poor training performance in terms of training speed.

\subsection{Layer-wise Partitioning and Merging}
\begin{figure*}[h!]
     \begin{subfigure}[b]{0.62\textwidth}
         \includegraphics[width=\textwidth]{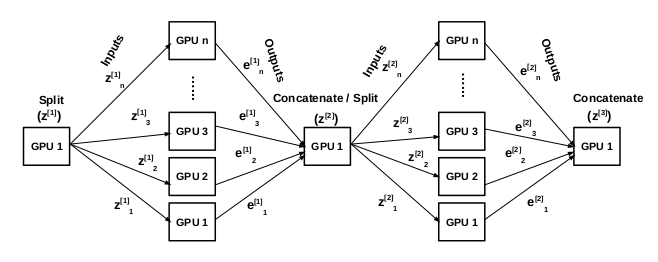}
         \caption{Layer partitioning}
         \label{fig1_1}
     \end{subfigure}
     \hfill
     \begin{subfigure}[b]{0.39\textwidth}
         \includegraphics[width=\textwidth]{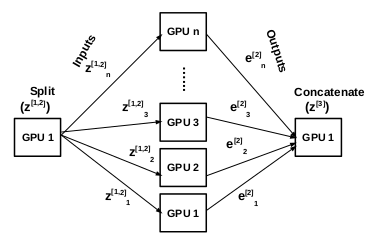}
         \caption{Layer merging}
         \label{fig1_2}
     \end{subfigure}
    
\caption{Layer partitioning and merging using $n$ number of GPUs. \label{fig0}}
\end{figure*}

To mitigate the aforementioned problem, we propose a novel layer-wise partitioning and merging method. The method uses layer-wise partitioning and merging operations to solve forward and backward locking problems by performing computations of network layers across available multiple devices rather than a single device as in the existing layer-wise partitioning methods. In the layer-wise partitioning operation, we  break the layers with multiple dimensions into different sub-layers, among the set of available devices $D$ (such as GPUs). For instance, we split a first DNN layer $p^{[1]}$ into $\lbrace p_{1}^{[1]}, p_{2}^{[1]},\ldots,p_{n}^{[1]} \rbrace$, where $n$ is the cardinality of the set of available devices $D$, where $D = {d_{i}, \,\,\ i \in [1,n]}$. Subsequently, we partition the new set of DNN layers into $Z$ sub-modules in which each sub-module comprises a stack either from similar or different layers. As shown in Fig. \ref{fig1_1}, each $d_{i}$ computes each partition $z^{j}_{i}$ of sub-module $j \in Z$, where $i \leq n$. The outputs are concatenated and re-partitioned for the next sub-module $j+1$. However, this method incurs high communication overhead due to frequent data movement among the devices. To address this challenge, we apply a merging operation where some sub-modules are merged as shown in Fig. \ref{fig1_2} such that the output of a sub-module are sent directly to the next sub-module without involving devices computation.\\
In addition, the sub-module $z$  performs a forward and a backward pass using activation input and gradient from module $z-1$, respectively. From equation~\ref{eqn1}, in the forward pass for at iteration $t$, the activation $a_{z-1}^{t}$ of $z-1$ is used as input to the sub-module $z$ and produces activation $a_{z}^{t}$. Likewise, from equation~\ref{eqn2}, in the backward pass at iteration $t$, the gradient from sub-module $z-1$ is fed input into sub-module $z$ to produce a new gradient.

\begin{figure*}[h!]
     \begin{subfigure}[b]{0.50\textwidth}
         \includegraphics[width=\textwidth]{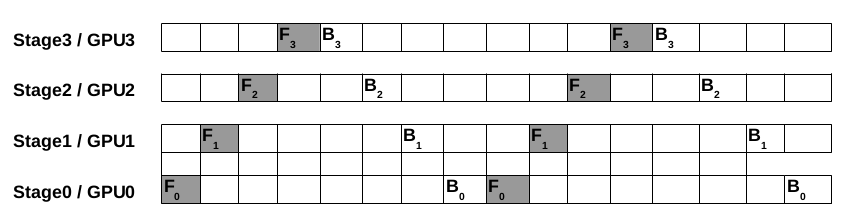}
         \caption{Model parallelism with non-parallel forward and backward pass computations}
         \label{pd:non-pip}
     \end{subfigure}
     \hfill
     \begin{subfigure}[b]{0.49\textwidth}
         \includegraphics[width=\textwidth]{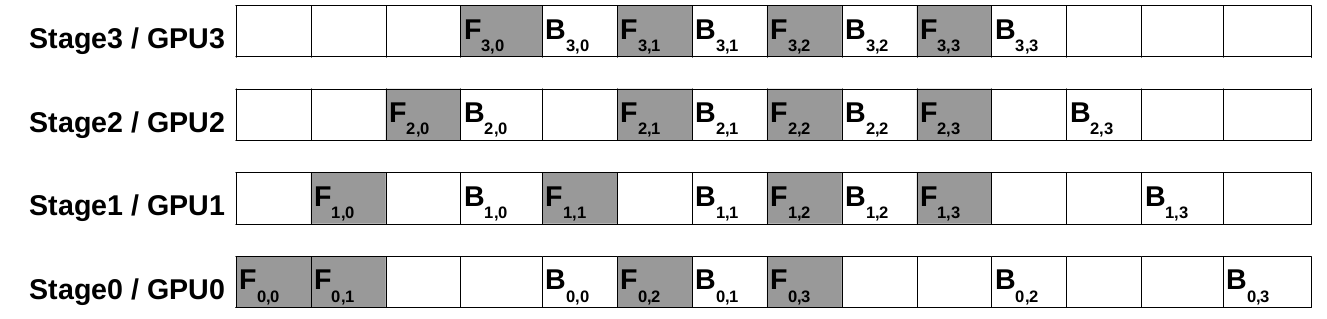}
         \caption{ Model parallelism with parallel forward and backward pass computations}
         \label{pd:pip2}
     \end{subfigure}
    
\caption{Forward and backward pass computations \label{fig1}}
\medskip
\small
The left figure is model parallelism with non-parallel forward and backward pass computations; a single data batch across GPUs leads to several unutilized hardware. The figure on the right shows parallel forward and backward pass computations with mini-batches for all GPUs.
\end{figure*}

\subsection{Forward Pass and Backward Pass Parallelization}
To further improve DNN training performance, we address the update locking problem by formulating a cost model and parallelizing the forward pass and backward pass to optimize the total execution cost.
\subsubsection{Cost Model}
We formulate a cost model to optimize the computation
time performance of the forward Pass and backward pass parallelization methods. The cost is defined as:

\begin{equation}
\min_{ F, B}T( F, B) =  \sum_{i,j=1}^{n} T(F_{i,j}) +\sum_{i,j=1}^{n} T(B_{i,j}) 
\end{equation}

where $T(F_{i,j})$ and $T(B_{i,j})$  are the cost functions of forward pass and backward pass computations respectively Thus, the goal of the cost model is to minimise the total execution cost $min(T)$.

\subsubsection{Forward Pass and Backward Pass Parallelisation and Optimisation}
Here, we parallelize forward pass, backward pass and parameter synchronization processes to minimize the total execution cost $min(T)$. The forward pass and backward pass parallelization methods address the update locking problem caused by the sequential execution of forward pass and backward pass computations. Fig \ref{fig1} represents model parallelism with non-parallel and parallel forward and backward pass computations. As shown in figure~\ref{pd:non-pip}, the forward pass computes the activation of modules $F_{i}$ from $i = 1$ to $n$ and the backward pass calculates the error gradients $B_{i}$ in reverse order from output layer to input layer, that is $i = n$ to $i = 1$ (as in \ref{eq2}), sequentially at iteration $t$. To reduce dependencies among the layers for both activation and gradient calclations, we parallelize the forward and backward passes by splitting the dataset by the number of available GPUs such that $X = \lbrace x_{1}, x_{2}, \ldots, x_{n} \rbrace$; each data batch is divided into minibatches of size $b/n$. Each module $z_{i}$, where $i = \lbrace 1, 2, \ldots,n \rbrace$ is fed with different dataset chunks. Each module computation runs on multiple GPUs concurrently to ensure better forward and backward pass throughput as well as hardware utilization efficiency. At each iteration, we first perform input and output computations of a module for the forward and backward pass before transferring dependencies between the two propagations. $B_{i,j}$ represents the backward pass of module $i$ and micro-batch $j$ while $F_{i,j}$ represents the forward pass of module $i$ and micro-batch $j$. For instance, in Fig. \ref{pd:pip2}, the output computations of $B_{0,0}$ finish before the input computations of $F_{0,2}$ in stage $0$. Similar operations are performed in stages $1$, $2$, and $3$.

To improve training performance, asynchronous parameter update and gradient accumulation methods are used to perform parameter updates among the module with all training batches rather than a single batch. At the training process's start, the first batch training samples are used for forward pass calculation, and backward pass starts with the same batch immediately forward pass finishes. Then the forward and backward pass computations perform for the subsequent batches. The training process continuously feeds new training batches for forward and backward pass
calculation tasks with different GPUs to ensure throughput and system utilization.

\section{Experimental Evaluation}\label{exp}
In this section, we carry out the experiments on the proposed method. Section \ref{exgo} stipulates the experiments goals. In section \ref{uscd}, we describe the use case; a DNN model and dataset used. Section \ref{eem} provides definitions of evaluation metrics; speedup, accuracy and training time. Section \ref{sc} provides the hardware and software settings of the experiments. In section \ref{erd}, we discuss the experimental results and compare the performance with the existing parallel methods.

\begin{table}[h!]
\caption{ ADNI database Descriptions}
\label{tab3:table0}
\begin{tabular}{p{0.5cm}>{\centering\arraybackslash}p{1.7cm}>{\centering\arraybackslash}p{1.6cm}>{\centering\arraybackslash}p{1.8cm}>{\centering\arraybackslash}p{1.2cm}}
\toprule
Class            & Number/Size & Gender (Male/Female) & Age (Mean/Std) & MMSE (Mean/Std) \\
\hline
AD & 389/1.4GB & 202/187 & 75.95/7.53 & 23.28/2.03    \\ 
pMCI  & 172/484MB          & 105/67 & 75.57/7.13 & 26.59/1.71    \\ 
sMCI & 232/649MB          & 155/77 & 75.71/7.87 & 27.27/1.78  \\ 
NC &400/2.4GB   & 202/198 & 76.02/5.18 & 29.10/1.01   \\ \bottomrule
\end{tabular}
\end{table}

\subsection{Experiment Goals}\label{exgo}
The goals of the experiments are to implement the proposed method as described in \ref{meth} and evaluate, through realistic use cases, its scalability and performance in a multi-GPU environment. We also compare the speedup and accuracy performance with some state-of-the-art methods.

\subsection{Use Case Description}\label{uscd}
We consider two real cases: complex and non-complex models, to adequately evaluate our proposed method's performance and robustness. Specifically, we adopt 3D-ResAttNet for Alzheimer's disease as a complex model and ResNet for Classification of CIFAR-10 Images as a non-complex model. 3D-ResAttNet is a 3-dimensional network which is deeper and has higher computation loads than the ResNet, a 2-dimensional deep learning network.

\subsubsection{Use Case 1: 3D-ResAttNet for Alzheimer's Disease}
We apply the proposed method on our previous non-parallel 3D-ResAttNet for automatic detection of the progression of AD and its Mild Cognitive Impairments (MCIs) - Normal cohort (NC), Progressive MCI (pMCI), and Stable MCI (sMCI) from sMRI scans \cite{Xingye}. The network consists of 3D Conv blocks, Residual self-attention blocks, and Explainable blocks. 3D convolutions exploit a 3D filter to calculate the low-level feature representations of the output shape as a 3-dimensional volume space. The residual self-attention block combines two important network layers: the residual network layer and the Self-attention layer. The residual network layer comprises two Conv3D blocks consisting of $3 \times 3$ 3D convolution layers, 3D batch normalization and Rectified Linear Unit (ReLU). The explainable block uses 3D Grad-CAM to improve the model decision. We adopt the dataset from the Alzheimer's Disease Neuroimaging Initiative (ADNI) database (\url{http://adni.loni.usc.edu}) as the benchmark for the performance evaluation. The dataset has four classes of MRI scans images, developed in 2003 by Dr Michael W.Weiner under the public-private partnership. As shown in Table \ref{tab3:table0}, it contains 1193 MRI scans 389 Alzheimer's Disease (AD), 400 Normal Cohort (NC), 232 Stable Mild Cognitive Impairment (sMCI) and 172 Progressive Mild Cognitive Impairment (pMCI) patients.

\subsubsection{Use Case 2: ResNet for Classification of CIFAR-10 Images} 
We also apply Residual Networks, ResNet18 \cite{SARWINDA2021423} and ResNet34 \cite{Tian2021AGP} for the classification of the CIFAR-10 images \cite{Giuste2020CIFAR10IC} to further evaluate the robustness of our proposed method. The networks consist of 2D convolutional layers with  $3 \times 3$ filters, batch normalization,  rectified linear unit and residual block layers. The network ends with an average pooling layer and a fully-connected layer. The CIFAR-10 dataset contains 60,000 images with $32 \times 32$ pixels, divided into 10000 test images and 50000 training images. The images are classified into ten classes - aeroplane, automobile, bird, cat, deer, dog, frog, horse, ship, and truck; each has 6,000 images. The CIFAR-10 dataset is commonly used for benchmarking DNN models.
\begin{figure*}[h!]
     \begin{subfigure}[b]{0.49\textwidth}
         \includegraphics[width=\textwidth]{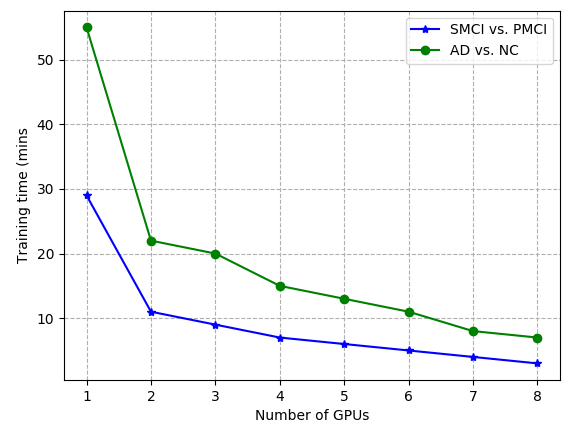}
         \caption{sMCI vs. pMCI and AD vs. NC classification tasks on 3D-ResAttNet18}
         \label{fig2_1}
     \end{subfigure}
     \hfill
     \begin{subfigure}[b]{0.49\textwidth}
         \includegraphics[width=\textwidth]{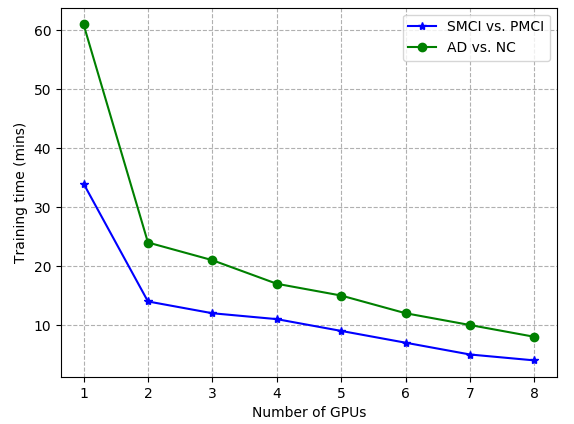}
         \caption{sMCI vs. pMCI and AD vs. NC classification tasks on 3D-ResAttNet34}
         \label{fig2_2}
     \end{subfigure}
    \hfill
     \begin{subfigure}[b]{0.49\textwidth}
         \includegraphics[width=\textwidth]{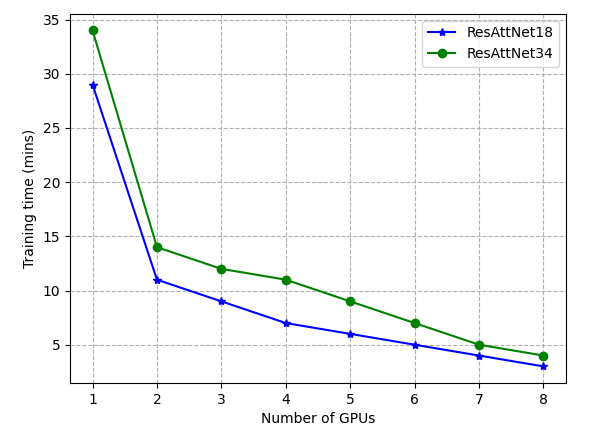}
         \caption{3D-ResAttNet18 vs. 3D-ResAttNet34 for sMCI vs. pMCI classification tasks}
         \label{fig2_3}
     \end{subfigure}
     \hfill
     \begin{subfigure}[b]{0.49\textwidth}
         \includegraphics[width=\textwidth]{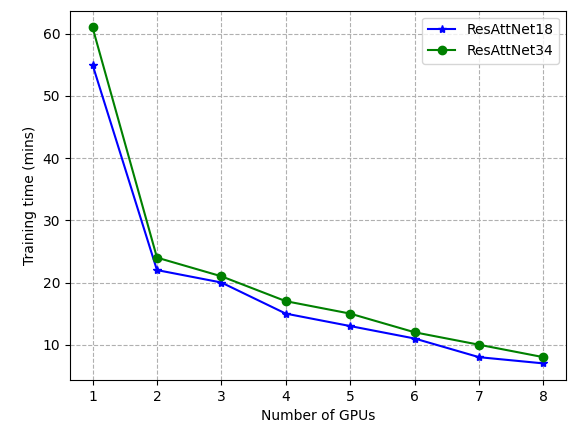}
         \caption{3D-ResAttNet18 vs. 3D-ResAttNet34 for AD vs. NC classification tasks}
         \label{fig2_4}
     \end{subfigure}
\caption{Use case 1: Training time performance of 3D-ResAttNet18 and 3D-ResAttNet34 using the proposed method \label{fig2}}
\end{figure*}

\begin{figure*}[h!]
     \begin{subfigure}[b]{0.49\textwidth}
         \includegraphics[width=\textwidth]{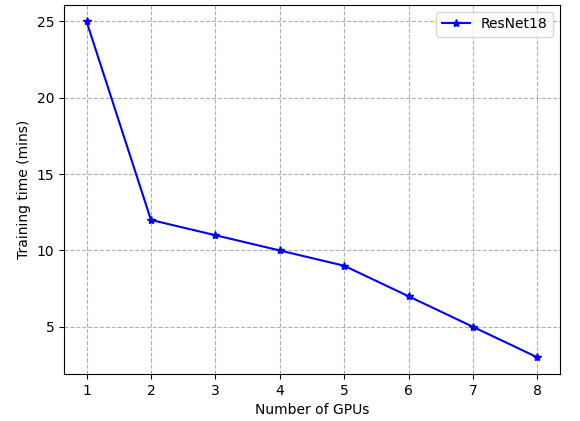}
         \caption{ResNet18}
         \label{fig8_1}
     \end{subfigure}
     \hfill
     \begin{subfigure}[b]{0.49\textwidth}
         \includegraphics[width=\textwidth]{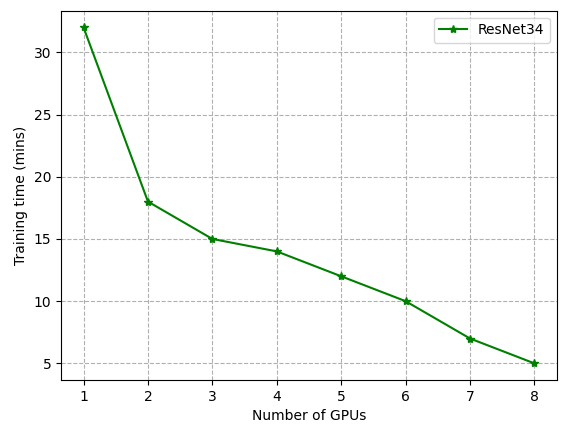}
         \caption{ResNet34}
         \label{fig8_2}
     \end{subfigure}
    
\caption{Use case 2: Training time performance of ResNet18 and ResNet34 using the proposed method \label{fig8}}
\end{figure*}

\subsection{Experimental Evaluation Metrics}\label{eem}
We adopted Speedup ($S$), Accuracy ($ACC$) and Training Time ($TT$) for performance evaluation of out proposed method. $S$ measures the scalability and computing performance and defined as:

\begin{equation}
S = T_{s}/T_{p}
\end{equation}

$T_{s}$ denotes  the computing time on a single GPUm and $T_{p}$ indicates the computing time on $p$ GPUs. $ACC$ measures the classification accuracy and is defined as:

\begin{equation}
ACC = (TP + TN)/(TP + TN + FP + FN) 
\end{equation}

where $TP$ = True positives, $FP$ = False positives, $TN$ = True negatives and $FN$ = False negatives. $TT$ measures time taken for training of the DNN models using the proposed approach and other existing parallel methods

\subsection{System Configuration}\label{sc}
\textbf{Hardware: } Our experiments are conducted on an Amazon Web Service (AWS) EC2 p3.16xlarge instance. The p3.16xlarge instance consists of 8 NVIDIA Tesla V100 GPUs with NVLink technology, 128GB GPU memory, 64 vCPUs, 4488GB memory, and 25Gbps network bandwidth.\\                                                                                                                                                                                                               
\textbf{Software: } we exploit the following software configuration and installation: Ubuntu 18.04 as the backbone for other software installation, Python 3.7.3, PyTorch 1.2.0 as the deep learning framework \cite{PyTorch}, Torchvision 0.4.0, Numpy 1.15.4, Tensorboardx 1.4, Matplotlib 3.0.1, Tqdm 4.39.0, nibabel, fastai, and NVIDIA Collective Communications Library (NCCL) CUDA toolkit 10.2 - a library of multi-GPU collective communication primitives \cite{Nvidia}.

\begin{table}[h!]
\begin{center}
\caption{Use case 1: training performances ResAttNet18 and ResAttNet34 with ADNI dataset using the proposed method} 
\label{tab3:table1}
\begin{tabular}{|>{\centering}p{0.7cm}|>{\centering}p{0.4cm}|>{\centering}p{0.4cm}|>{\centering}p{0.4cm}|>{\centering}p{0.4cm}|>{\centering}p{0.4cm}|>{\centering}p{0.4cm}|>{\centering}p{0.4cm}|>{\centering}p{0.4cm}|>{\centering}p{0.4cm}|>{\centering}p{0.4cm}|>{\centering}p{0.4cm}|>{\centering}p{0.4cm}|>{\centering}p{0.4cm}|>{\centering}p{0.5cm}|} \hline
&\multicolumn{4}{ c| }{3D-ResAttNet18} &  \multicolumn{4}{ c| }{3D-ResAttNet34} \\      \cline{2-9}
 &\multicolumn{2}{ c| }{sMCI vs. pMCI} &  \multicolumn{2}{ c| }{AD vs. NC}&  \multicolumn{2}{ c| }{sMCI vs. pMCI}&  \multicolumn{2}{ c| }{AD vs. NC} \\ \cline{2-9}
\#GPUs &ACC&TT (mins)& ACC&TT (mins)& ACC&TT (mins)& ACC&TT (mins)
 \tabularnewline \hline
 1 & 0.79 & 29 & 0.93 & 55 & 0.81 & 34 & 0.94 & 61\tabularnewline \hline
 2 & 0.80 & 11 & 0.92 & 22 & 0.81 & 14 & 0.94 & 24\tabularnewline \hline
 3 & 0.81 & 9 & 0.92 & 20 & 0.82 & 12 & 0.95 & 21\tabularnewline \hline
 4 & 0.80 & 7 & 0.93 & 15 & 0.83 & 11 & 0.93 & 17\tabularnewline \hline
 5 & 0.80 & 6 & 0.94 & 13 & 0.81 & 9 & 0.93 & 15\tabularnewline \hline
 6 & 0.79 & 5 & 0.93 & 11 & 0.82 & 7 & 0.95 & 12\tabularnewline \hline
 7 & 0.79 & 4 & 0.93 & 8 & 0.83 & 5 & 0.94 & 10\tabularnewline \hline
 8 & 0.80 & 3 & 0.93 & 7 & 0.82 & 4 & 0.94 & 8 \tabularnewline \hline

\end{tabular}
\end{center}
\end{table}

\begin{table}[h!]
\begin{center}
\caption{Use case 2: training performances ResNet18 and ResNet34 with CIFAR-10 using the proposed method} 
\label{tab3:table2}
\begin{tabular}{|>{\centering}p{0.7cm}|>{\centering}p{0.4cm}|>{\centering}p{0.4cm}|>{\centering}p{0.4cm}|>{\centering}p{0.4cm}|>{\centering}p{0.4cm}|>{\centering}p{0.4cm}|>{\centering}p{0.4cm}|} \hline
&\multicolumn{2}{ c| }{ResNet18} &  \multicolumn{2}{ c| }{ResNet34} \\      \cline{2-5}
 
\#GPUs &ACC&TT (mins)& ACC&TT (mins)
 \tabularnewline \hline
 1 & 0.93 & 25 & 0.94 & 32 \tabularnewline \hline
 2 & 0.94 & 12 & 0.94 & 18 \tabularnewline \hline
 3 & 0.94 & 11 & 0.93 & 15 \tabularnewline \hline
 4 & 0.93 & 10 & 0.94 & 14 \tabularnewline \hline
 5 & 0.94 & 9 & 0.94 & 12 \tabularnewline \hline
 6 & 0.93 & 7 & 0.94& 10 \tabularnewline \hline
 7 & 0.93 & 5 & 0.94 & 7 \tabularnewline \hline
 8 & 0.94 & 3 & 0.94 & 5  \tabularnewline \hline

\end{tabular}
\end{center}
\end{table}

\begin{figure*}[h!]
     \begin{subfigure}[b]{0.47\textwidth}
         \includegraphics[width=\textwidth]{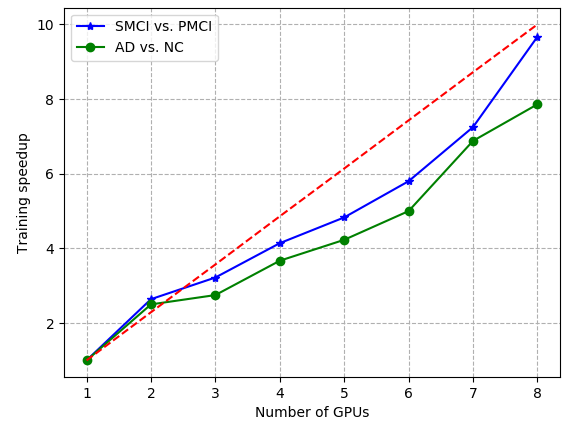}
         \caption{sMCI vs. pMCI and AD vs. NC classification tasks on 3D-ResAttNet18}
         \label{fig3_1}
     \end{subfigure}
     \hfill
     \begin{subfigure}[b]{0.47\textwidth}
         \includegraphics[width=\textwidth]{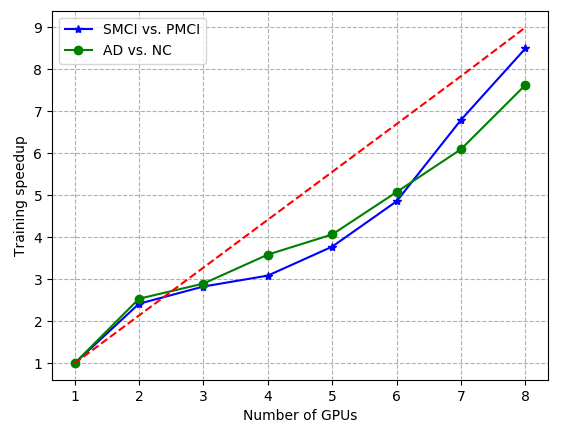}
         \caption{sMCI vs. pMCI and AD vs. NC classification tasks on 3D-ResAttNet34}
         \label{fig3_2}
     \end{subfigure}
    \hfill
     \begin{subfigure}[b]{0.47\textwidth}
         \includegraphics[width=\textwidth]{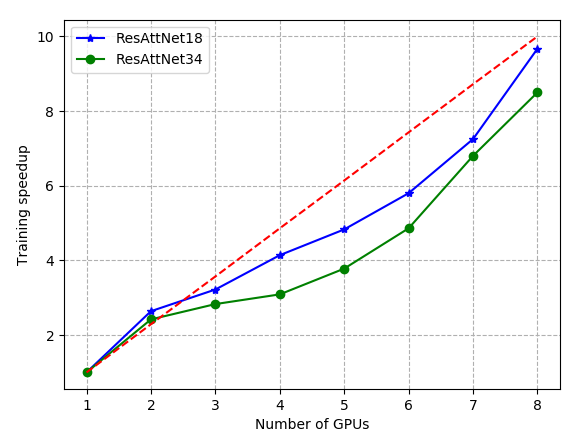}
         \caption{3D-ResAttNet18 vs. 3D-ResAttNet34 for sMCI vs. pMCI classification tasks}
         \label{fig3_3}
     \end{subfigure}
     \hfill
     \begin{subfigure}[b]{0.47\textwidth}
         \includegraphics[width=\textwidth]{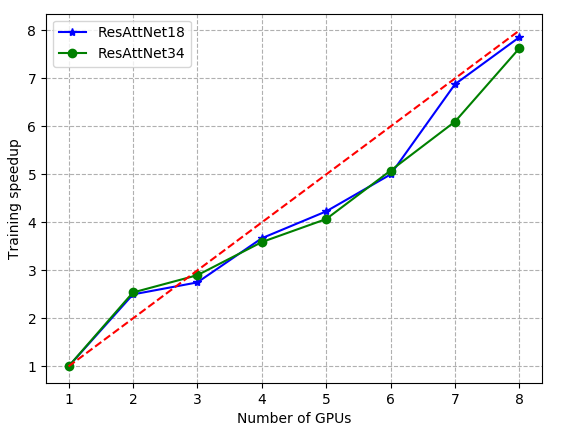}
         \caption{3D-ResAttNet18 vs. 3D-ResAttNet34 for AD vs. NC classification tasks}
         \label{fig3_4}
     \end{subfigure}
\caption{Use case 1: speedup performance of 3D-ResAttNets using the proposed method \label{fig3}}
\end{figure*}

\begin{figure*}[h!]
     \begin{subfigure}[b]{0.49\textwidth}
         \includegraphics[width=\textwidth]{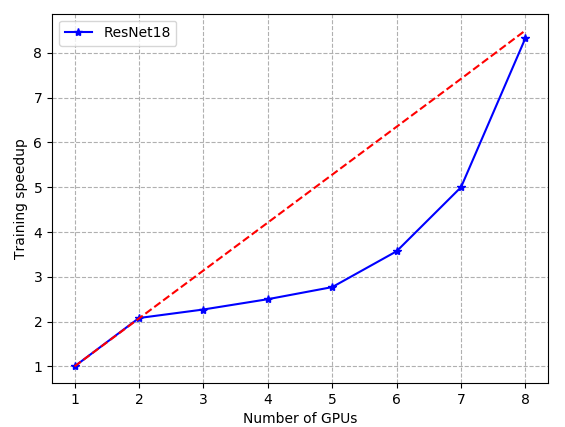}
         \caption{ResNet18}
         \label{fig7_1}
     \end{subfigure}
     \hfill
     \begin{subfigure}[b]{0.49\textwidth}
         \includegraphics[width=\textwidth]{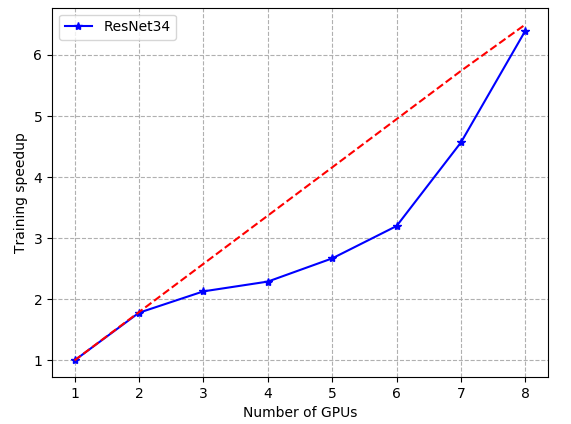}
         \caption{Reset34}
         \label{fig7_2}
     \end{subfigure}
    
\caption{Use case 2: speedup performance of ResNet18 and ResNet34 using the proposed method  \label{fig7}}
\end{figure*}

\begin{figure*}[h!]
     \begin{subfigure}[b]{0.47\textwidth}
         \includegraphics[width=\textwidth]{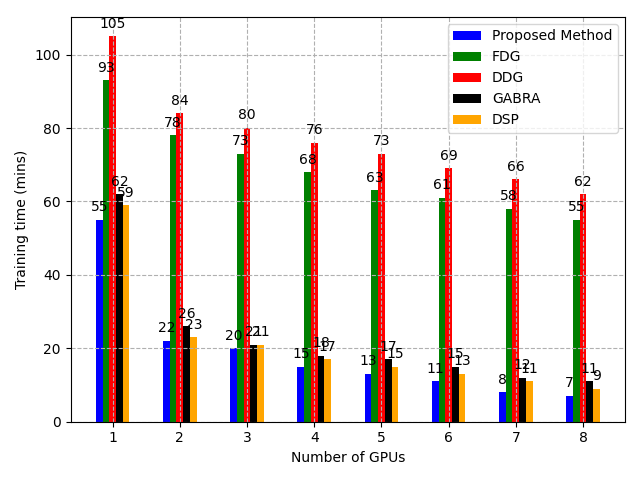}
         \caption{Training Time of 3D-ResAttNet18 over AD vs. NC Dataset}
         \label{fig5_5}
     \end{subfigure}
     \hfill
     \begin{subfigure}[b]{0.47\textwidth}
         \includegraphics[width=\textwidth]{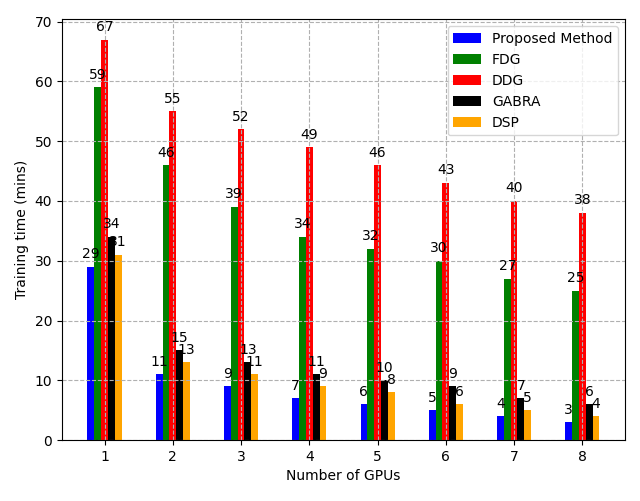}
         \caption{Training Time of 3D-ResAttNet18 over sMCI vs. pMCI Dataset}
         \label{fig5_6}
     \end{subfigure}
    \hfill
     \begin{subfigure}[b]{0.47\textwidth}
         \includegraphics[width=\textwidth]{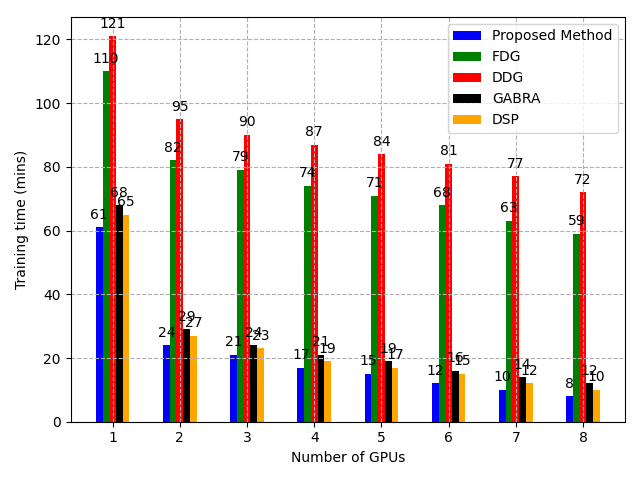}
         \caption{Training Time of 3D-ResAttNet34 over AD vs. NC Dataset}
         \label{fig5_7}
     \end{subfigure}
     \hfill
     \begin{subfigure}[b]{0.47\textwidth}
         \includegraphics[width=\textwidth]{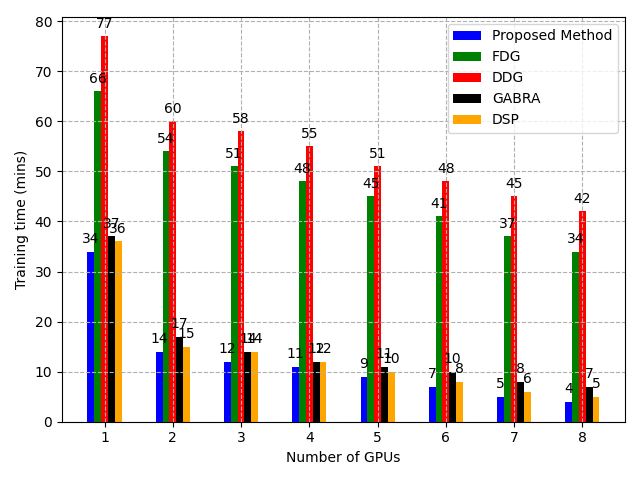}
         \caption{Training Time of 3D-ResAttNet34 over sMCI vs. pMCI Dataset}
         \label{fig5_8}
     \end{subfigure}
\caption{Use case 1: training time of 3D-ResAttNet18 and 3D-ResAttNet34 using the proposed method, FDG, DDG, DSP and GABRA \label{fig5}}
\end{figure*}

\begin{figure*}[h!]
     \begin{subfigure}[b]{0.49\textwidth}
         \includegraphics[width=\textwidth]{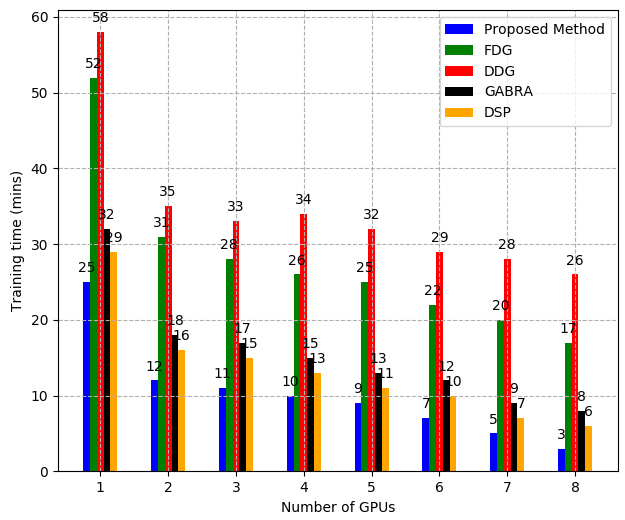}
         \caption{ResNet18}
         \label{fig6_1}
     \end{subfigure}
     \hfill
     \begin{subfigure}[b]{0.49\textwidth}
         \includegraphics[width=\textwidth]{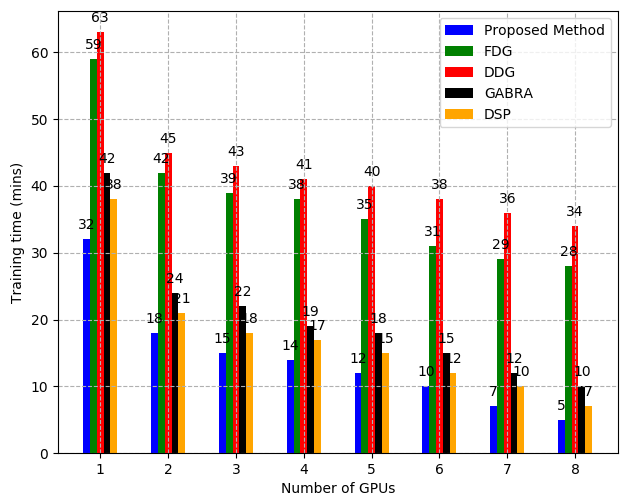}
         \caption{Reset34}
         \label{fig6_2}
     \end{subfigure}
    
\caption{Use case 2: training time of ResNet18 and ResNet34 with CIFAR-10 using the proposed method, FDG, DDG, DSP and GABRA \label{fig6}}
\end{figure*}

\subsection{Experiments Results and Discussions}\label{erd}
We investigated the accuracy and speedup performances of the proposed method on the use case 1, 3D-ResAttNets for two classification tasks: sMCI vs pMCI and AD vs NC, and use case 2, ResNets on CIFAR-10 with varying numbers of GPUs (ranging from 1 to 8). Furthermore, we compare performances of the proposed against  GABRA \cite{Akintoye2021AHP}, our previous parallelisation method, and other three state-of-art methods, including DDG, DSP and FDG. We used Rectified linear unit(Relu) as the activation function and optimized model parameters with SGD, a stochastic optimization algorithm. In addition, we set other training parameters, including a batch size of six samples, cross-entropy as the loss function, and 50 epochs for better convergence. We set initial learning rate (LR) as $1 \times 10^{-4}$, then reduced by $1 \times 10^{-2}$ with increased iterations.

\subsubsection{Training Time}
Tables \ref{tab3:table1} and \ref{tab3:table2} show the training results of use case 1, 3D-ResAttNets on the two classification tasks: sMCI vs pMCI and AD vs NC, on 3D-ResAttNet18 and 3D-ResAttNet34 models, and use case 2, ResNets with CIFAR-10 using the proposed parallelisation method. Figures \ref{fig2} and \ref{fig8} visualise the performance of use cases 1 and 2 using the proposed parallelisation method in terms of the training time with the varying number of GPUs, respectively. Both show that as the number of GPUs increases, the training time decreases. For instance, in use case 1, the sMCI vs pMCI classification task on 3D-ResAttNet18 gives 33 mins when using a single GPU, 14 mins with 2 GPUs, and 5 mins with 8 GPUs. The same performance trend is seen for AD vs NC classification tasks on 3D-ResAttNet18 and sMCI vs pMCI and AD vs NC classification tasks on 3D-ResAttNet18. Likewise, in use case 2, the training of ResNet18 with CIFAR-10 images gives 25 mins with a single GPU, mins with 2 GPUs, and 3 mins with 3 GPUs. ResNet34 with CIFAR-10 dataset gives 32 mins on a single GPU, 18 mins with 2 GPUs, and 5 mins with 3 GPUs.

\subsubsection{Speedup}
We investigate the relationship between speedup and the number of GPUs. The speedup measures the scalability and computing performance and can be defined as $T_{s}/T_{p}$, where $T_{s}$ represents training time on a single GPU and $T_{p}$ denotes training time on multiple GPUs. Figures \ref{fig3} (c) and (d) show, for use case 1, the speedup performance of ResAttNets using the proposed method calculated based on the training time with varying number of GPUs. The figure shows that the speedup increases linearly with the number of GPUs, which illustrates the scalability of the proposed method. Specifically, for  AD vs. NC classification task with 3D-ResAttNet18, the training speedup on 1, 2, 3, 4, 5, 6, 7, and 8 GPUs are 1, 2.5, 2.75, 3.67, 4.23, 5, 6.88 and 7.86 respectively. The sMCI vs pMCI classification task with 3D-ResAttNet18 the training speedup for 1, 2 ,3, 4, 5, 6, 7, and 8 GPUs are 1, 2.64, 3.22, 4.14, 4.83, 5.8, 7.25 and 9.67. A similar speedup performance trend was also observed in the sMCI vs pMCI and AD vs NC classification tasks with 3D-ResAttNet34.\\

Figure \ref{fig7} shows, for use case 2, the speedup performance of ResNets - ResNet18 and ResNet34 using the proposed method, also showing that the speedup increases linearly with the number of GPUs.

\subsubsection{Accuracy}
Tables~\label{tab3:table1} and \label{tab3:table2} show the behaviours of the test accuracy of the image classification tasks on use case 1, 3D-ResAttNet18 and 3D-ResAttNet34 models, and use case 2, ResNet18 and ResNet34 using the proposed parallelisation method respectively. They show no specific pattern of performance for the test accuracy with different numbers of GPUs. Specifically, there is no significant relationship between the test accuracy and the number of GPUs. For instance, in use case 1, the proposed method provides test accuracies:
 0.94, 0.94, 0.95, 0.93, 0.93, 0.95, 0.94 and 0.94  on 1, 2 ,3, 4, 5, 6, 7, and 8 GPUs respectively using 3D-ResAttNet34. The same behaviours are also shown in the accuracy performance of 3D-ResAttNet18 and use case 2,  ResNet18 and ResNet34 with the different numbers of GPUs. The proposed method does not provide a significant deviation from non-parallel training accuracies.

\begin{table}[h!]
\begin{center}
\caption{Comparison of the memory usage of the existing methods and our proposed method. The proposed method took an average of 66\% and 54\% of the memory to train 3D-ResAttNet34 and ResNet34 models, respectively and outperforms the GABRA, DDG, DSP and FDG.} 
\label{tab7:table1}
\begin{tabular}{|>{\centering}p{2.1cm}|>{\centering}p{2.5cm}|>{\centering}p{2.5cm}|} \hline
 Methods&3D-ResAttNet34 (AD vs. NC) &  ResNet34 (CIFAR-10)\tabularnewline \hline
Proposed method & 66\% & 54\%  \tabularnewline \hline
FDG & 78\% & 71\%   \tabularnewline \hline
DDG &89\% & 79\%   \tabularnewline \hline
DSP & 69\% & 60\%  \tabularnewline \hline
GABRA &73\% & 65\%   \tabularnewline \hline

\end{tabular}
\end{center}
\end{table}

\subsubsection{Comparison of the Proposed Method and Existing Methods}
We have compared the proposed method with our previous parallelisation method, GABRA and other three state-of-art methods: DDG, DSP and FDG. Figures \ref{fig5} and \ref{fig6} show the experiment results of use cases 1 and 2, respectively. The proposed method outperforms the GABRA, DDG, DSP and FDG in terms of training time. For instance, in use case 1, for 3D-ResAttNet18 on AD vs NC and sMCI vs pMCI classification tasks, the training time incurred by the proposed method is lower than GABRA, DDG, DSP and FDG.  Similarly, there are related trends when comparing the proposed method with the GABRA, DDG, DSP and FDG for the distributed training of 3D-ResAttNet34 for two classification tasks: AD vs NC and sMCI vs pMCI, and use case 2, Resnet18 and ResNet34 with CIFAR-10 dataset.  \\
In addition, Table \ref{tab7:table1} shows the comparison of the memory usage of the exiting methods and our proposed method. Our proposed method consumed an average of 66\% and 54\% of the system memory to train 3D-ResAttNet34 and ResNet34 models, respectively. In all results, our proposed method outperforms the existing methods, FDG DDG, DSP and GABRA, in terms of memory usage.

\section{Conclusion}\label{pd:conc}
In this paper, we have proposed layer-wise partitioning and merging to solve forward and backward locking problems by performing computations of network layers across available multiple devices rather than a single device. We have also proposed a forward pass and backward pass parallelisation method to address the update locking problem associated with the sequential execution of forward pass and backward pass computations. We applied the proposed method to train two CNNs - our previous 3D Residual Attention Deep Neural Network (3D-ResAttNet) model on real-world Alzheimer's Disease (AD) datasets and the Residual Network (ResNet) model CIFAR-10 dataset. The experimental results show that the proposed method achieves linear speedup, demonstrating its scalability and efficient computing capability. In addition, the comparison evaluation shows that the proposed method can earn a considerable speedup and reduced memory consumption when training the deep feed-forward neural network and outperforms the three state-of-art methods on the same benchmark datasets.
\section*{Acknowledgment}
The work reported in this paper has formed part of the project by Royal Society - Academy of Medical Sciences Newton Advanced Fellowship (NAF \textbackslash R1\textbackslash 180371).

\bibliographystyle{abbrv}
\bibliography{bare_jrnl_Layer-wise-Lloyd-08-07-2022}

\begin{thebibliography}{10}

\bibitem{dl:Abadi}
M.~Abadi, P.~Barham, J.~Chen, Z.~Chen, A.~Davis, J.~Dean, M.~Devin,
  S.~Ghemawat, G.~Irving, M.~Isard, and et~al.
\newblock Tensorflow: A system for large-scale machine learning.
\newblock In {\em Proceedings of the 12th USENIX Conference on Operating
  Systems Design and Implementation}, OSDI'16, pages 265--283, USA, 2016.
  USENIX Association.

\bibitem{Akintoye2021AHP}
S.~B. Akintoye, L.~Han, X.~Zhang, H.~Chen, and D.~Zhang.
\newblock A hybrid parallelization approach for distributed and scalable deep
  learning.
\newblock {\em ArXiv}, abs/2104.05035, 2021.

\bibitem{dl:Eugene}
E.~Belilovsky, M.~Eickenberg, and E.~Oyallon.
\newblock Decoupled greedy learning of {CNN}s.
\newblock In {\em Proceedings of the 37th International Conference on Machine
  Learning}, volume 119 of {\em Proceedings of Machine Learning Research},
  pages 736--745. PMLR, 13--18 Jul 2020.

\bibitem{Boehm1}
M.~Boehm, S.~Tatikonda, B.~Reinwald, P.~Sen, Y.~Tian, D.~R. Burdick, and
  S.~Vaithyanathan.
\newblock Hybrid parallelization strategies for large-scale machine learning in
  systemml.
\newblock {\em Proc. VLDB Endow.}, 7(7):553--564, 2014.

\bibitem{Cheng2021DerivationOT}
Y.~Cheng.
\newblock Derivation of the backpropagation algorithm based on derivative
  amplification coefficients.
\newblock {\em ArXiv}, abs/2102.04320, 2021.

\bibitem{DBLP:DaiZDZX19}
W.~Dai, Y.~Zhou, N.~Dong, H.~Zhang, and E.~P. Xing.
\newblock Toward understanding the impact of staleness in distributed machine
  learning.
\newblock In {\em 7th International Conference on Learning Representations,
  ICLR, 2019, New Orleans, LA, USA, May 6-9, 2019}. OpenReview.net, 2019.

\bibitem{dp:Dean}
J.~Dean, G.~S. Corrado, R.~Monga, K.~Chen, M.~Devin, Q.~V. Le, M.~Z. Mao,
  M.~Ran-zato, A.~Senior, P.~Tucker, K.~Yang, , and A.~Y. Ng.
\newblock Large scale distributed deep networks.
\newblock In {\em Proceedings of NIPS}, pages 1232--1240, 2012.

\bibitem{Dryden2019}
N.~Dryden, N.~Maruyama, T.~Moon, T.~Benson, M.~Snir, and B.~Van~Essen.
\newblock Channel and filter parallelism for large-scale cnn training.
\newblock In {\em Proceedings of the International Conference for High
  Performance Computing, Networking, Storage and Analysis}, SC '19, New York,
  NY, USA, 2019. Association for Computing Machinery.

\bibitem{Fan2018GeneralBA}
F.~Fan, W.~Cong, and G.~Wang.
\newblock General backpropagation algorithm for training second-order neural
  networks.
\newblock {\em International journal for numerical methods in biomedical
  engineering}, 34 5(5):e2956, 2018.

\bibitem{Geng2019}
J.~Geng, D.~Li, and S.~Wang.
\newblock Elasticpipe: An efficient and dynamic model-parallel solution to dnn
  training.
\newblock {\em Proceedings of the 10th Workshop on Scientific Cloud Computing},
  2019.

\bibitem{Giuste2020CIFAR10IC}
F.~O. Giuste and J.~C. Vizcarra.
\newblock Cifar-10 image classification using feature ensembles.
\newblock {\em ArXiv}, abs/2002.03846, 2020.

\bibitem{Gower2019SGDGA}
R.~M. Gower, N.~Loizou, X.~Qian, A.~Sailanbayev, E.~Shulgin, and
  P.~Richt{\'a}rik.
\newblock Sgd: General analysis and improved rates.
\newblock {\em ArXiv}, abs/1901.09401, 2019.

\bibitem{Hu2018}
J.~Hu, L.~Shen, and G.~Sun.
\newblock Squeeze-and-excitation networks.
\newblock {\em 2018 IEEE/CVF Conference on Computer Vision and Pattern
  Recognition}, pages 7132--7141, 2018.

\bibitem{Huang2019}
Y.~Huang, Y.~Cheng, D.~Chen, H.~Lee, J.~Ngiam, Q.~V. Le, and Z.~Chen.
\newblock Gpipe: Efficient training of giant neural networks using pipeline
  parallelism.
\newblock {\em ArXiv}, abs/1811.06965, 2019.

\bibitem{Huo2018TrainingNN}
Z.~Huo, B.~Gu, and H.~Huang.
\newblock Training neural networks using features replay.
\newblock {\em ArXiv}, abs/1807.04511, 2018.

\bibitem{dl:dg2}
Z.~Huo, B.~Gu, Q.~Yang, and H.~Huang.
\newblock Decoupled parallel backpropagation with convergence guarantee.
\newblock In {\em ICML}, 2018.

\bibitem{dl:Zhouyuan}
Z.~Huo, B.~Gu, Q.~Yang, and H.~Huang.
\newblock Decoupled parallel backpropagation with convergence guarantee.
\newblock In {\em ICML}, 2018.

\bibitem{dl:Max}
M.~Jaderberg, W.~M. Czarnecki, S.~Osindero, O.~Vinyals, A.~Graves, D.~Silver,
  and K.~Kavukcuoglu.
\newblock Decoupled neural interfaces using synthetic gradients.
\newblock In {\em Proceedings of the 34th International Conference on Machine
  Learning - Volume 70}, ICML'17, pages 1627--1635. JMLR.org, 2017.

\bibitem{Jia2018ExploringHD}
Z.~Jia, S.~Lin, C.~Qi, and A.~Aiken.
\newblock Exploring hidden dimensions in parallelizing convolutional neural
  networks.
\newblock {\em ArXiv}, abs/1802.04924, 2018.

\bibitem{Jia2019}
Z.~Jia, M.~Zaharia, and A.~Aiken.
\newblock Beyond data and model parallelism for deep neural networks.
\newblock {\em ArXiv}, abs/1807.05358, 2019.

\bibitem{Kingma2015AdamAM}
D.~P. Kingma and J.~Ba.
\newblock Adam: A method for stochastic optimization.
\newblock {\em ArXiv}, abs/1412.6980, 2015.

\bibitem{Ko2018EdgeHostPO}
J.~Ko, T.~Na, M.~Amir, and S.~Mukhopadhyay.
\newblock Edge-host partitioning of deep neural networks with feature space
  encoding for resource-constrained internet-of-things platforms.
\newblock {\em 2018 15th IEEE International Conference on Advanced Video and
  Signal Based Surveillance (AVSS)}, pages 1--6, 2018.

\bibitem{dp:Alex2}
A.~Krizhevsky.
\newblock One weird trick for parallelizing convolutional neural networks.
\newblock {\em ArXiv}, abs/1404.5997, 2014.

\bibitem{Li2020}
A.~H. Li and A.~Sethy.
\newblock Semi-supervised learning for text classification by layer
  partitioning.
\newblock {\em ICASSP 2020 - 2020 IEEE International Conference on Acoustics,
  Speech and Signal Processing (ICASSP)}, pages 6164--6168, 2020.

\bibitem{Liu2020GenerativeFR}
X.~Liu, C.~Wu, M.~Menta, L.~Herranz, B.~Raducanu, A.~D. Bagdanov, S.~Jui, and
  J.~van~de Weijer.
\newblock Generative feature replay for class-incremental learning.
\newblock {\em 2020 IEEE/CVF Conference on Computer Vision and Pattern
  Recognition Workshops (CVPRW)}, pages 915--924, 2020.

\bibitem{8593683}
Y.~Liu, A.~Xu, and Z.~Chen.
\newblock Map-based deep imitation learning for obstacle avoidance.
\newblock In {\em 2018 IEEE/RSJ International Conference on Intelligent Robots
  and Systems (IROS)}, pages 8644--8649, 2018.

\bibitem{Martins2}
F.~Martins Campos~de Oliveira and E.~Borin.
\newblock Partitioning convolutional neural networks for inference on
  constrained internet-of-things devices.
\newblock 09 2018.

\bibitem{Martins1}
F.~Martins Campos~de Oliveira and E.~Borin.
\newblock Partitioning convolutional neural networks to maximize the inference
  rate on constrained iot devices.
\newblock {\em Future Internet}, 11(10), 2019.

\bibitem{Deepak2019}
D.~Narayanan, A.~Harlap, A.~Phanishayee, V.~Seshadri, N.~R. Devanur, G.~R.
  Ganger, P.~B. Gibbons, and M.~Zaharia.
\newblock Pipedream: Generalized pipeline parallelism for dnn training.
\newblock In {\em Proceedings of the 27th ACM Symposium on Operating Systems
  Principles}, SOSP '19, pages 1--15, New York, NY, USA, 2019. Association for
  Computing Machinery.

\bibitem{Nvidia}
Nvidia.
\newblock Nccl,
  https://docs.nvidia.com/deeplearning/nccl/install-guide/index.html, 2021.

\bibitem{Ono2019HybridDP}
J.~Ono, M.~Utiyama, and E.~Sumita.
\newblock Hybrid data-model parallel training for sequence-to-sequence
  recurrent neural network machine translation.
\newblock In {\em PSLT@MTSummit}, 2019.

\bibitem{PyTorch}
PyTorch.
\newblock Pytorch-deep learning framework that puts python first, 2020.

\bibitem{9039657}
D.~Saguil and A.~Azim.
\newblock A layer-partitioning approach for faster execution of neural
  network-based embedded applications in edge networks.
\newblock {\em IEEE Access}, 8:59456--59469, 2020.

\bibitem{SARWINDA2021423}
D.~Sarwinda, R.~H. Paradisa, A.~Bustamam, and P.~Anggia.
\newblock Deep learning in image classification using residual network (resnet)
  variants for detection of colorectal cancer.
\newblock {\em Procedia Computer Science}, 179:423--431, 2021.
\newblock 5th International Conference on Computer Science and Computational
  Intelligence 2020.

\bibitem{6853593}
F.~Seide, H.~Fu, J.~Droppo, G.~Li, and D.~Yu.
\newblock On parallelizability of stochastic gradient descent for speech dnns.
\newblock In {\em 2014 IEEE International Conference on Acoustics, Speech and
  Signal Processing (ICASSP)}, pages 235--239, 2014.

\bibitem{8675232}
L.~Song, J.~Mao, Y.~Zhuo, X.~Qian, H.~Li, and Y.~Chen.
\newblock Hypar: Towards hybrid parallelism for deep learning accelerator
  array.
\newblock In {\em 2019 IEEE International Symposium on High Performance
  Computer Architecture (HPCA)}, pages 56--68, 2019.

\bibitem{Szegedy2016}
C.~Szegedy, V.~Vanhoucke, S.~Ioffe, J.~Shlens, and Z.~Wojna.
\newblock Rethinking the inception architecture for computer vision.
\newblock {\em 2016 IEEE Conference on Computer Vision and Pattern Recognition
  (CVPR)}, pages 2818--2826, 2016.

\bibitem{Tian2021AGP}
J.~Tian, D.~Yung, Y.-C. Hsu, and Z.~Kira.
\newblock A geometric perspective towards neural calibration via sensitivity
  decomposition.
\newblock {\em ArXiv}, abs/2110.14577, 2021.

\bibitem{dl:Wu6}
Y.~Wu, M.~Schuster, Z.~Chen, Q.~V. Le, M.~Norouzi, W.~Macherey, M.~Krikun,
  Y.~Cao, Q.~Gao, K.~Macherey, J.~Klingner, A.~Shah, M.~Johnson, X.~Liu,
  L.~Kaiser, S.~Gouws, Y.~Kato, T.~Kudo, H.~Kazawa, K.~Stevens, G.~Kurian,
  N.~Patil, W.~Wang, C.~Young, J.~R. Smith, J.~Riesa, A.~Rudnick, O.~Vinyals,
  G.~Corrado, M.~Hughes, and J.~Dean.
\newblock Google's neural machine translation system: Bridging the gap between
  human and machine translation.
\newblock {\em ArXiv}, abs/1609.08144, 2016.

\bibitem{9156835}
A.~Xu, Z.~Huo, and H.~Huang.
\newblock On the acceleration of deep learning model parallelism with
  staleness.
\newblock In {\em 2020 IEEE/CVF Conference on Computer Vision and Pattern
  Recognition (CVPR)}, pages 2085--2094, 2020.

\bibitem{Yang2019}
B.~Yang, J.~Zhang, J.~Li, C.~R{\'e}, C.~R. Aberger, and C.~D. Sa.
\newblock Pipemare: Asynchronous pipeline parallel dnn training.
\newblock {\em ArXiv}, abs/1910.05124, 2019.

\bibitem{Zhan2019}
J.~Zhan and J.~Zhang.
\newblock Pipe-torch: Pipeline-based distributed deep learning in a gpu cluster
  with heterogeneous networking.
\newblock {\em 2019 Seventh International Conference on Advanced Cloud and Big
  Data (CBD)}, pages 55--60, 2019.

\bibitem{Xingye}
X.~Zhang, L.~Han, W.~Zhu, L.~Sun, and D.~Zhang.
\newblock An explainable 3d residual self-attention deep neural network for
  joint atrophy localization and alzheimer's disease diagnosis using structural
  mri.
\newblock {\em IEEE journal of biomedical and health informatics}, PP, 2021.

\bibitem{234801}
L.~Zhou, H.~Wen, R.~Teodorescu, and D.~H. Du.
\newblock Distributing deep neural networks with containerized partitions at
  the edge.
\newblock In {\em 2nd {USENIX} Workshop on Hot Topics in Edge Computing
  (HotEdge 19)}, Renton, WA, July 2019. {USENIX} Association.

\bibitem{dl:dg1}
H.~Zhuang, Y.~Wang, Q.~Liu, and Z.~Lin.
\newblock Fully decoupled neural network learning using delayed gradients.
\newblock {\em IEEE transactions on neural networks and learning systems}, PP,
  2021.

\bibitem{dl:Huiping}
H.~Zhuang, Y.~Wang, Q.~Liu, and Z.~Lin.
\newblock Fully decoupled neural network learning using delayed gradients.
\newblock {\em IEEE transactions on neural networks and learning systems}, PP,
  2021.

\end{thebibliography}

\ifCLASSOPTIONcaptionsoff
  \newpage
\fi

\ifCLASSOPTIONcaptionsoff
  \newpage
\fi

\begin{IEEEbiography}
[{\includegraphics[width=1in,height=1.25in,clip,keepaspectratio]{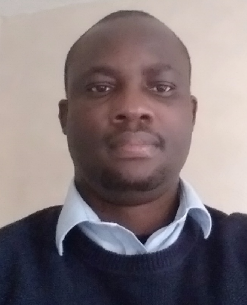}}]{Samson B. Akintoye}
received the Ph.D. degree in Computer Science from University of the Western Cape, South Africa, 2019. He is currently working as a research associate in the Department of Computing and Mathematics, Manchester Metropolitan University, United Kingdom. His current research interests include parallel and distributed computing, deep learning, and cloud computing.
\end{IEEEbiography}

\begin{IEEEbiography}
[{\includegraphics[width=1in,height=1.25in,clip,keepaspectratio]{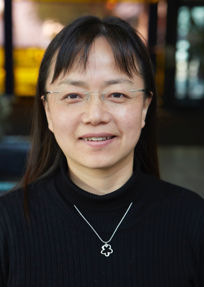}}]{Liangxiu Han}
received the Ph.D. degree in computer science from Fudan University, Shanghai, China, in 2002. She is currently a Professor of computer science with Department of Computing and Mathematics, Manchester Metropolitan University. Her research areas mainly lie in the development of novel big data analytics and development of novel intelligent architectures that facilitates big data analytics (e.g., parallel and distributed computing, Cloud/Service-oriented computing/ data intensive computing) as well as applications in different domains using various large datasets (biomedical images, environmental sensor, network traffic data, web documents, etc.). She is currently a Principal Investigator or Co-PI on a number of research projects in the research areas mentioned above.
\end{IEEEbiography}

\begin{IEEEbiography}
    [{\includegraphics[width=1in,height=1.25in,clip,keepaspectratio]{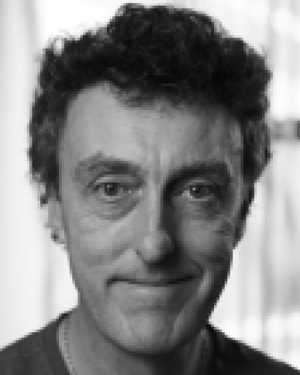}}]{Huw Lloyd}
    received the BSc degree in Physics from Imperial College, London, UK and the PhD degree in Astrophysics from the University of Manchester, UK. He is currently a Senior Lecturer with the Department of Computing and Mathematics, Manchester Metropolitan University. His research interests cover a range of theoretical and applied topics in machine learning, evolutionary computation, combinatorial and continuous optimization.
\end{IEEEbiography}

\begin{IEEEbiography}
[{\includegraphics[width=1in,height=1.25in,clip,keepaspectratio]{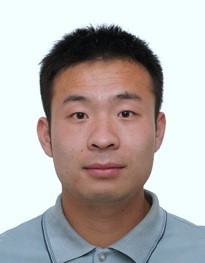}}]{Xin Zhang}
is associate researcher in Manchester Metropolitan University (MMU), he received the B.S degree from The PLA Academy of Communication and Commanding, China, in 2009 and Ph.D. degree in Cartography and Geographic Information System from Beijing Normal University(BNU), China, in 2014. His current research interests include remote sensing image processing and deep learning.
\end{IEEEbiography}

\begin{IEEEbiography}
[{\includegraphics[width=1in,height=1.25in,clip,keepaspectratio]{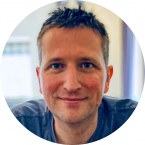}}]{Darren Dancey}
is the Head of the Department for Computing and Mathematics. He has a research background in Artificial Intelligence with a PhD in Artificial Neural Networks. His teaching interests centre on Software Development. He has recently taught courses in Data Structures and Algorithms, Comparative Programming languages and applied courses such as Website Development and Mobile Application Development. In recent years, he has concentrated on creating collaborations and knowledge exchange between universities and industry with a focus on the SME sector. He has led several large projects funded by Innovate UK, the Digital R\&D Fund for the Arts, and the European Research Council. He is on the organising committee for the Manchester Raspberry Pi Jam and sits on the BCS Manchester branch committee.
\end{IEEEbiography}

\begin{IEEEbiography}
[{\includegraphics[width=1in,height=1.25in,clip,keepaspectratio]{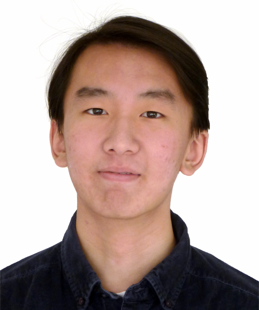}}]{Haoming Chen}
 is studying for a master’s degree in Computer Science and Artificial Intelligence in University of Sheffield. His current research interests include machine learning and Artificial Intelligence.
\end{IEEEbiography}

\begin{IEEEbiography}
[{\includegraphics[width=1in,height=1.25in,clip,keepaspectratio]{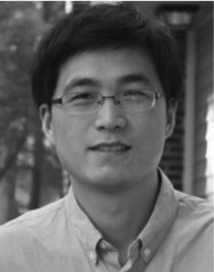}}]{Daoqiang Zhang}
 received the B.Sc. and Ph.D. degrees in computer science from Nanjing University of Aeronautics and Astronautics, Nanjing, China, in 1999 and 2004, respectively. He is currently a Professor in the Department of Computer Science and Engineering, Nanjing University of Aeronautics and Astronautics. His current research interests include machine learning, pattern recognition, and biomedical image analysis. In these areas, he has authored or coauthored more than 100 technical papers in the refereed international journals and conference proceedings.
\end{IEEEbiography}





\end{document}